\documentclass[12pt]{iopart}
\usepackage{iopams}
\usepackage{epsfig}
\def\({\left(}
\def\){\right)}
\def\[{\left[}
\def\]{\right]}
\def\be{\begin{eqnarray}}
\def\ee{\end{eqnarray}}
\def\ne{\nonumber\end{eqnarray}}
\def\b{\beta}

\def\e{\varepsilon}

\def\exp#1{\mathop{\rm exp}\nolimits \left\{#1\right\}}

\def\b{\beta}

\def\para#1{\vskip0.2cm\noindent{\bf #1}}
\begin{document}

\title[Exponentially localized solutions]
{Exponentially localized solutions of the Klein-Gordon equation}

\author{M. V. Perel  and I. V. Fialkovsky}
\address{Department of Theoretical Physics,  State University of
Saint-Petersburg, Russia} \ead{ifialk@gmail.com}

\begin{abstract}
Exponentially localized solutions of the Klein-Gordon equation for
two and three space variables are presented. The solutions depend
on four free parameters. For some relations between the
parameters, the solutions describe wave packets filled with
oscillations whose amplitudes decrease in the Gaussian way with
distance from a point running with group velocity along a straight line. The
solutions are constructed using exact complex solutions of the
eikonal equation and may be regarded as ray solutions with
amplitudes involving one term. It is also shown that the
multidimensional nonlinear Klein-Gordon equation can be reduced to
an ordinary differential equation with respect to the complex
eikonal.

\noindent\textit{Translated from Zapiski Nauchnykh Seminarov POMI, Vol. 275, 2001, pp. 187--198.}
\end{abstract}

\para{Introduction.}
The construction of various highly localized solutions of the wave
equation
\be
\phi_{tt} -\phi_{xx}-\phi_{yy} -\phi_{zz}= 0
\ee
is a subject of several publications (see [1-8]).
In particular, in [1] a solution that is
exponentially localized in the vicinity of a point running with
velocity of light is given. In this paper, we generalize the
result of [6] by presenting a family of localized solutions that
includes the one given in [6] as a special case.

Based on the
results obtained for the wave equation, we construct a family of
particle-like solutions of the Klein-Gordon equation
\be
h^2 (u_{tt}- u_{xx} -u_{yy}) + u = 0,\quad h= const.
\ee
These solutions have finite energy
and describe wave packets with central frequency $\omega$ and wave number
$k$, where $\omega^2 = k^2 + 1/h^2$. Their amplitudes decrease exponentially
with distance from a point running along a straight line with
group velocity $v = d\omega/dk$.
By analogy with the solutions of the
wave equation, we call them \textit{Gaussian wave packets}.

All the solutions of the Klein-Gordon equation from this class can
be represented in the form
$$
u = Af(iS/h),
$$
where the
function $S$ satisfies the eikonal (or Hamilton-Jacobi) equation
\be
S^2_t - S^2_x - S^2_y = 1,
\ee
the amplitude factor $A$ does
not depend on the coordinates, $S$ and $A$ are independent on $h$,
and $f$ is expressed in terms of the Hankel function.

The comparison of our results with
those available in the literature shows that one of the
solutions that we found for the three-dimensional Klein-Gordon
equation coincides with the solution given in [7]. We also
consider in detail another solution from the constructed family,
which depends only on one variable $S$ that is one of the exact
complex solutions of the eikonal equation. It turned out that the
search for a solution dependent only on $S$ of the nonlinear
Klein-Gordon equation is reduced to the solution of an ordinary
differential equation. This is also true for nonlinear
Klein-Gordon equations with arbitrary number of space variables.

In constructing particle-like solutions of the wave equation, we
followed the idea by Ziolkowski [3] and sought such solutions in
the form of a superposition of Gaussian beams, which are solutions
localized near rays. The latter solutions were found for the first
time in papers of Brittingham [1] and Kiselev [2] and belong to
the class of relatively undistorted waves, in the terminology of
Courant and Hilbert [10]. The construction of the solution of the
Klein-Gordon equation employs the simple observation that, taking
the Fourier transform of a solution of wave equation (1), say with
respect to $z$,
\be
u(x, y, t) = \int_{-\infty}^\infty dz \phi(x, y, z, t) e^{iz/h},
\ee
we obtain a solution of the Klein-Gordon equation (2).

\para{Generalized Gaussian packets for the wave equation.}
 We start from
the solution of wave equation (1), described by the formula
\be
    \phi_b(x, y, z, t, q)
        = \frac{\exp{iq \Theta_1}}{(\b-i\e_1)^{1/2}(\b-i\e_2)^{1/2}}
\ee
(see [1, 2, 6]), where the notation
\be
    \Theta_1 = x - t + \frac{y^2}{\b-i\e_1} + \frac{z^2}{\b-i\e_2},\quad \b= x + t,
\ee
is introduced. The function $\phi_b$ satisfies (1) for any $q$,
$\e_1$, and $\e_2$, and in the case of $\e_1 > 0$, $\e_2> 0$, and
$q > 0$ it is a Gaussian beam, which means that it is
localized in the Gaussian way in the vicinity of the $x$ axis. We
seek particle-like solutions of Eq. (1) in the form of a
superposition of Gaussian beams:
\be
\phi^{(\nu)}_p (x, y, z, t)
    = \int_0^\infty dq     F^{(\nu)}(q) \phi_b(x, y, z, t, q), (7)
\ee
 where $F^{(\nu)}(q)$ is a particular
function depending on the parameter $\nu$. We put
\be
    F^{(\nu)}(q) \equiv a q^{-\nu-1} e^{-\e(q+\sigma^2/q)}, (8)
\ee
where $\nu$, $\sigma$, and $\e$ are arbitrary constants, $\sigma > 0$,
$\e > 0$, and $a = (4\e\sigma^2)^\nu/(2\sqrt\pi)$.
It can easily be shown that (7) is
reduced to an integral representation of the Hankel function $H^{(1)}_\nu$
of the first kind [12] and
\be
\phi^{(\nu)}_p (x, y, z, t)
    = C \frac{s^\nu H^{(1)}_\nu (s)}{\sqrt{(\b-i\e_1) (\b-i\e_2)}},
    s= 2i\sigma\e \(1 - \frac{i\Theta_1}{\e}\)^{1/2} ,
\ee
where $C = i 2^{\nu-1}\sqrt{\pi}$.
It is worth noting that $s$ satisfies the Hamilton-Jacobi
equation $s^2_t = s^2_x+s^2_y +s^2_z$ for wave equation (1). We note that
for $\nu = 1/2$, formula (9) yields a solution of the wave equation
presented earlier in [6]:
\be
\phi^({1/2)}_p (x, y, z, t)
    = \frac{\exp{-2\sigma\e \sqrt{1-   i\Theta_1/\e}}}{\sqrt{(\b-i\e_1) (\b-i\e_2)}}.
\ee
This solution depends on four
free parameters $\e$, $\e_1$, $\e_2$, and $\sigma$. It is established in [6] that if
all these parameters are positive, it is localized in the Gaussian
way near the point $x = y = 0$ and $z = ct$ that runs with velocity of
light $c = 1$ along the $x$ axis. The asymptotics of the solutions of
(1) of the form (9) with respect to the large argument have the
same exponential factor as (10). Arguments similar to those
adduced in [6] prove their localization. Therefore, (9) is a
localized solution generalizing (10).

\para{Gaussian beams for the Klein-Gordon equation.}
Here we give a solution $u_b$ of the
Klein-Gordon equation, which has a Gaussian localization near a
ray, in order to use it in construction of
particle-like solutions of the Klein-Gordon equation. To find such
a $u_b$, we calculate the Fourier transform (4) with respect to $z$ of
expression (5):
\be
u_b(x, y, t; q) = \frac{\sqrt{\pi} e^{i\pi/4-\e_2/(4qh2)} }{\sqrt{q}}
    \frac{\exp{i\Theta q - i\b/(4qh^2)}}{\sqrt{\b -i\e_1}} ,
\ee
 where
\be
 \Theta  =x - t + \frac{y^2}{\b - i\e_1}.
\ee
This solution was found first in [8]. We call such solutions of
the Klein-Gordon equation \textit{Gaussian beams}, by analogy with
solutions of wave equations.

\para{Gaussian packets for the Klein-Gordon equation.}
Taking the Fourier transform (4) with
respect to $z$ of the both sides of Eq. (7), we obtain solutions of
the Klein-Gordon equation in the form of an expansion in the beam
solutions $u_b$ from (11):
\be
u^{(\nu)}_p (x, y, t) = \int_0^\infty dq F^{(\nu)}(q) u_b(x,y, t; q).
\ee
Substituting $u_b$ of the form (11) and $F^{(\nu)}(q)$ of the
form (8) into (13), we arrive at
\be
u^{(\nu)}_p (x, y, t) = C_1
    \frac{S^{\nu+1/2}_p H^{(1)}_{\nu+1/2}(S_p/h)}
        {(\b - i\e_1)^{1/2}(\b - 4i\e\sigma^2 h^2 - i\e_2)^{\nu+1/2}}
\ee
with
\be
    S_p = i\[(\Theta + i\e)(\b - 4i\e\sigma^2 h^2 - i\e_2)\]^1/2 .
\ee
The function $S_p$ does
not depend on $\nu$. Formula (14) yields a family of exact solutions
of the Klein-Gordon equation, which are all particle-like as will
be seen in what follows. This family depends on four free parameters
$\nu$, $\e$, $\e_1$, and $\sigma$. In what follows we use a different and simpler
parametrization of (14). The constant factor in (14) is equal to
$C_1=\pi(8\e\sigma^2h^2+2\e_2)^{\nu-1/2}h^{\nu+3/2}e^{i\pi\pi(\nu+1)}/\sqrt2$.

\para{The ray interpretation of Gaussian beams and Gaussian packets.}
The introduction of a new
free parameter $\kappa$ by the relation $\kappa = qh$ enables us to rewrite (11)
in the form of a ray series, which is reduced to only one term
\be
    u_b = c_b \frac{\exp{iS_b/h}}{(\b - i\e_1)^{1/2}},\quad c_b = const,
\ee
with
\be
    S_b = \kappa\Theta -\frac{\b}{4\kappa}.
\ee
Here, the complex phase function $S_b$ is independent of $h$
and satisfies eikonal equation (3). We also note that $u_b$ may be
regarded as a reference solution for the asymptotic construction
presented by Maslov in [11].

The solutions $u^{(\nu)}_p$ can conveniently
be written in terms of a new free parameter
$\gamma = 4\e\sigma^2h^2 + \e_2$
instead of $\sigma$ and $\e_2$ and $\mu = \nu + 1/2$ instead of $\nu$. Now,
\be
u^{(\mu-1/2)}_p
    = C_1 \frac{ S^\mu_p H^{(1)}_\mu (S_p/h)}{ (\b - i\e_1)^{1/2}(\b - i\gamma)^\mu} .
\ee
The complex
phase function
\be
    S_p = i \[(\Theta + i\e)(\b - i\gamma)\]^{1/2}
\ee
 is also
independent of $h$ and satisfies eikonal equation (3). In the case
of a half-integer $\mu$, the Hankel functions are reduced to
elementary functions. For the values $\mu = 1/2$ and $\mu = -1/2$, we have
\be
    u^{(0)}_p = C_2\frac{\exp{iS_p/h}}{(\b - i\e_1)^{1/2}(\b -i\gamma)^{1/2}}
\ne
and
\be
    u^{(-1)}_p = C_3\frac{\exp{iS_p/h}}{S_p} \sqrt{\frac{\b - i\gamma}{\b - i\e_1}} ,
\ee
respectively, where $C_2$ and
$C_3$ are constants. These solutions, as well as (16), are ray
solutions reduced to their zero-order term.

\para{Properties of the solutions.}
 First we show that the solution $u_b$ in (16) is localized
near the $x$ axis. Separating the real and imaginary parts in the
complex phase function $S_b$ from (17) and introducing the notation
\be
\tilde\kappa = \(\kappa -\frac1{ 4\kappa}\),\quad
    \tilde\omega= \sqrt{\tilde\kappa^2 + 1},\quad
    \Delta_y = \sqrt{2\frac{\e^2_1 + \b^2}{\e_1(\tilde\kappa + \tilde\omega)}}, (21)
\ee
we obtain
\be
iS_b = i(\tilde\kappa x -\tilde\omega t) -\frac{y^2}{\Delta^2_y}
    +i\frac{y^2} {\Delta^2_y} \cdot\frac{ \b}{ \e_1 }.
\ee
For $|\b| \ll \e_1$, the solution $u_b$ describes a wave with frequency
$\omega = \tilde\omega/h$ and wave number $k = \tilde\kappa/h$, which propagates along the $x$
axis and decays in the Gaussian way with distance from the axis.
The degree of localization near the x axis is determined by the
parameter $h\Delta_y$. The solution shows the Gaussian localization near
the $x$ axis for any finite values of $\b$, and the degree of
localization decreases as $\b$ grows.

Now we describe the asymptotic
behavior of $u^{(\mu-1/2)}_p$. As has already been mentioned, all the
functions $S_p$ are independent of $\mu$. The asymptotic expressions of
the Hankel function for large values of $|S_p|/h$ differ only in the
phase factor, and all solutions of our family behave as
\be
u^{(\mu-1/2)}_p (x, y, t) \sim
    C_3\frac{S^{\mu-1/2}_p \exp{iS_p/h} }{(\b - i\e_1)^{1/2}(\b - i\gamma)^\mu}
\ee
with $C_3 = C_1\sqrt{2/\pi}\,e^{-i\pi\mu/2-i\pi/4}$.

Now we investigate the dependence of
$S_p$ on the space variables and time. Separating the real and
imaginary parts in the subradical expression from (19), we obtain
\be
\hskip-2cm
iS_p =
    -\[
    \gamma\e + x^2 - t^2 + y^2\frac{\b^2 + \gamma\e_1}{\b^2 + \e_1^2}
     +2i\(
        \sqrt{\gamma\e}(\tilde\omega t -\tilde\kappa x) +
        y^2\frac{\b(\e_1 - \gamma)}{\b^2 + \e_1^2}
        \)
    \]^{1/2},
\ee
where $\tilde\kappa$ and $\tilde\omega$ are defined by
(21) with
$\kappa = \sqrt\gamma/4\e$. We now prove a rough estimate valid for any
value of $t$, which implies that any solution from the family
described by (14) has finite energy. First we note the simple fact
that for real $a_1$ and $a_2$,
${\rm Re}\sqrt{a_1 + ia_2} = \((a_1 +\sqrt{a^2_1 + a_2^2})/2\)^{1/2}$,
whence
$$
{\rm Re}\sqrt{a_1 + ia_2} \geq a_1\ {\rm for}\ a_1 \geq 0
$$
and thus
\be
|{\rm Re}(iS_p)| \geq \[
    \gamma\e + x^2
    - t^2 + y^2\frac{\b^2 + \gamma\e_1}{\b^2 + \e^2_1}
    \]^{1/2}.
\ee
Estimate (25) shows that
for a fixed time and sufficiently large values of the space
coordinates, the asymptotic expression for the Hankel functions is
applicable and the behavior of solution (18) is described by
formula (23). Therefore, the absolute value of any solution
$|u^{(\mu-1/2)}_p |$ decreases exponentially with a rise in the
coordinates, and therefore all these solutions have finite energy.
This estimate is valid for all values of the space coordinates in
case of $|t|\leq\gamma\e$.

Now we prove that, under certain restrictions
on $x$, $y$, and $t$ and some conditions on their parameters, the
solutions $u^{(\mu-1/2)}_p$ describe wave packets localized in the
Gaussian way. We expand $iS_p$ in a series, assuming that all
subradical terms dependent on $x$, $y$, and $t$ are small in comparison
with $\e\gamma$.
Keeping only the terms linear and quadratic in $x$, $y$, and
$t$, we find
\be
iS_p \sim i(\tilde\kappa x -\tilde\omega t) -\sqrt{\gamma\e}
    -\frac{(x - vt)^2}{\Delta^2_x} -\frac{y^2}{\Delta^2_y},
\ee
where $v$ stands for the group velocity
$v = d\tilde\omega/d\tilde\kappa = \tilde\kappa/\tilde\omega$ and
\be
\Delta_x = \frac{\sqrt{2\kappa\e}}{\tilde\omega},\quad
    \Delta_y = \sqrt{\frac{\e_1}{\kappa}}.
\ee
We recall that $\kappa = \sqrt\gamma/4\e$.

Under
the condition $\sqrt{\gamma\e} = 2\kappa\e \gg h$, the asymptotic form of the Hankel
function can be used in (18) for all values of $x$, $y$, and $t$.
Formula (26) can be applied for not-too-large $x$, $y$, and $t$.
However, for some values of the parameters $\kappa$, $\e$, and $\e_1$, the
domain of its applicability can be larger that the widths $\sqrt{h}\Delta_x$ and
$\sqrt{h}\Delta_y$ of the packet. In this domain, all the solutions from the
family $u^{(\mu-1/2)}_p$ demonstrate Gaussian localization near a
point running along the $x$ axis with group velocity $v$. Such a
behavior may occur for a certain relation between all the
parameters, which we write below in an interesting special case.

On the Fig. 1. the real part of $u^{(0)}_p$ given by (20) is presented
in conventional units: $t = 0$ (left),
 $t = 10$ (right); in both cases, $h = 0.3$, $\e = 10$, and $\kappa = 1$.

\begin{figure}
$$\begin{array}{cc}
  \epsfig{file=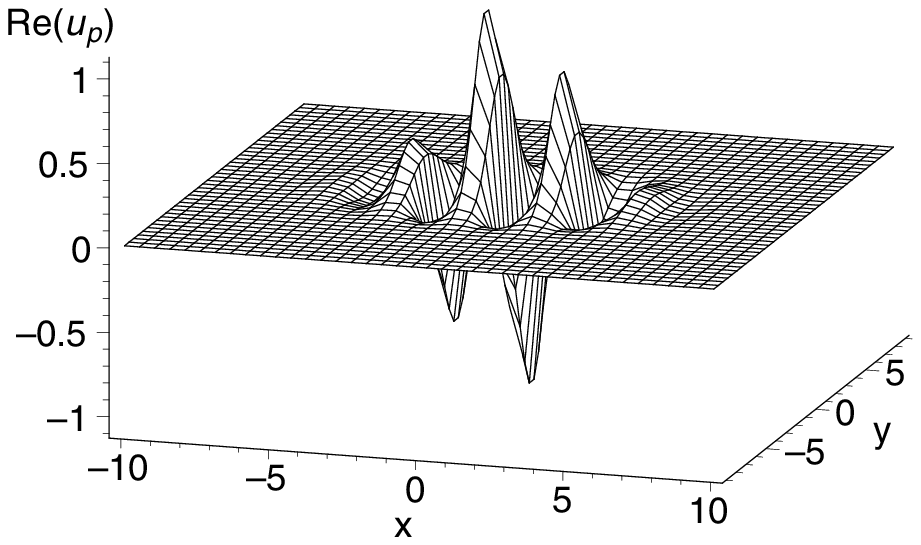,width=0.45\hsize}
 & \epsfig{file=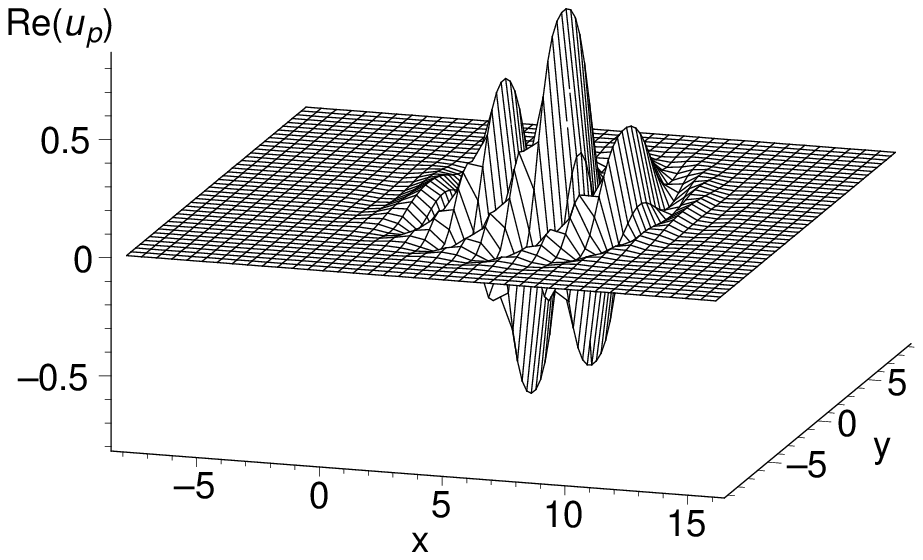,width=0.45\hsize}
 \\
\end{array}
$$
\caption{Real part of $u^{(0)}_p$
in conventional units: $t = 0$ (left),
 $t = 10$ (right).}
\end{figure}

\para{A simple example of a Gaussian packet.}
An interesting solution of the
Klein-Gordon equation (2) is obtained from (1) by setting
$\e_1 = \gamma\equiv 4\kappa^2\e$.
The solution $u^{(-1)}_p$ depends only on one variable $S_p$, and,
up to a constant factor, we have
\be
    u^{(-1)}_p = C_p \frac{e^{iS_p/h}}{ S_p} .
\ee
In
this case, the phase $S_p$ is given by
\be
S_p = i\sqrt{ b^2 + x^2 + y^2 - t^2 +2ib(\tilde\omega t - \tilde\kappa x)}
    \nonumber\\
   \qquad = i\sqrt{ (x - ib\tilde\kappa)^2 + y^2 - (t - ib\tilde\omega )^2}
\ee
with $ b = \sqrt{\gamma\e} \equiv 2\kappa\e$.
Here, $\tilde\kappa$ can be expressed in terms of $\kappa$, see (21).
However, it is more convenient to regard $\tilde\kappa$ and $b$ as independent
parameters. Then, $\tilde\omega  = \sqrt{\tilde\kappa^2 + 1}$.

The expansion of $S_p$ takes the
form (26) with $\Delta_x = \sqrt {b}/\tilde\omega$  and $\Delta_y = \sqrt{2b}$.
We describe the conditions
for its validity. The expansion of (29) is justified for $x\ll b$,
$y\ll  b$, $t \ll  b$, $t \ll  b/\tilde\omega$ ,
and $x \ll  b/\tilde\kappa$. Since $\tilde\omega  \geq 1$, the conditions
$t \ll  b/\tilde\omega$  and
$x \ll  \min(b, b/\tilde\kappa)$ are more restrictive. It can be seen
from (29) that the solution is localized in the Gaussian way near
a point running with group velocity if the transverse and
longitudinal widths $\sqrt{h}\Delta_x$ and $\sqrt{h}\Delta_y$, as well as the time width
$\sqrt{h}\Delta_x/v$, are all far less that the domain where the root in (29)
can be expanded, namely, for $\sqrt{h b} \tilde\omega  \ll  \min(b, b/\tilde\kappa)$,
$\sqrt{2hb} \ll  b$, and $\sqrt{hb}/\tilde\kappa \ll  b/\tilde\omega$.
Here, the most restrictive condition is
\be
    h \ll  b \frac{\tilde\kappa^2}{ \tilde\omega^2}.
\ee

\para{The case of three space variables.} All of the above
considerations can be generalized to three or more space
variables. Here we present the results for the three-dimensional
case and compare them with the results found in [7]. We turn to
the equation
\be
    h^2 (u_{tt} - u_{xx} - u_{yy} - u_{zz}) + u = 0,\quad h= const.
\ee
The construction of its localized solutions is based on the
expressions for localized solutions of the four dimensional wave
equation
\be
    \phi_{tt} - \phi_{xx} - \phi_{yy} - \phi_{zz} - \phi_{z_1z_1} = 0.
\ee
Axially symmetric
Gaussian beams for this equation are described by the expression
\be
\phi_b(x, y, z, z_1, t, q)
    = \frac{\exp{iq\Theta_1}}{ (\b - i\e_1)(\b - i\e_2)^{1/2}}.
\ee
In contrast to (6), here
\be
\Theta_1 = x - t + \frac{y^2}{\b - i\e_1} +\frac{z^2}{\b - i\e_1}
     + \frac{z^2_1}{\b - i\e_2},\quad \b= x + t.
\ee
We construct generalized Gaussian
packets for Eq. (32) by a formula similar to (7), starting from
the Gaussian beams and using weight function (8) in the same way
as we did this previously. Then we take the Fourier transform
\be
    u(x, y, z, t) = \int_{-\infty}^\infty dz_1 \phi (x, y, z, z_1, t) e^{iz_1/h}
\ee
of the both
sides of a formula analogous to (7). In doing so, we find
solutions of the Klein-Gordon equation (31) describing the
Gaussian particles
\be
    u^{(\mu-1/2)}_p = C_4\frac{ S^\mu_p H^{(1)}_\mu (S_p/h)}{ (\b- i\e_1)(\b - i\gamma)^\mu},
\ee
where $C_4 = const$ and $S_p$ is defined by (19) in which
\be
    \Theta = x - t + \frac{y^2}{\b - i\e_1} + \frac{z^2}{ \b - i\e_1} .
\ee
The Gaussian particles
$u^{(\mu-1/2)}_p$ in the $3$D case differ from the corresponding solutions
in the $2$D case (18) by a power of $\b - i\e_1$ in the denominator and
by the additional term involving the variable $z$ in phase (37). The
solution
\be
u^{(-1)}_p = C_5\frac{(\b - i\gamma)^{1/2}}{(\b -i\e_1)S_p}\exp{iS_p/h}
        = C_5 \frac{\exp{iS_p/h}}{(\b - i\e_1)(\Theta+i\e)^{1/2}}
\ee
with $C_5 = const$ was found earlier
in [7]. This is a Gaussian particle from (36) for $\mu = -1/2$. In the
three-dimensional case, as well as in the two-dimensional case, a
solution dependent only on the variables $S_p$ exists. This is
$u^{(-3/2)}_p$, where we must set $\e_1=\gamma$:
\be
     u^{(-3/2)}_p = C_6 \frac{H^{(1)}_1 (iS_p/h)}{ S_p} ,
\ee
where $C_6 = const$.

\para{On nonlinear Klein-Gordon equations.}
The above construction can be used in solving nonlinear Klein-
Gordon equations
\be
    h^2 (u_{tt} - u_{xx} - u_{yy}) + \phi(u) = 0
\ee
in the
two-dimensional case and
\be
    h^2 (u_{tt} - u_{xx} - u_{yy}-u_{zz}) + \phi (u) = 0
\ee
in the three-dimensional case. Seeking solutions in the form
$u = F(s)$, where $s = S_p/h$, we arrive at nonlinear differential
equations
\be
    F_{ss} + \frac2s F_s + \phi (F) = 0
\ee
and
\be
    F_{ss} + \frac3s F_{s} + \phi (F) = 0
\ee
in the case of (40) and (41), respectively.

\para{Acknowledgments.}
We are indebted to Professor A. P. Kiselev for his interest in this research and
his stimulating discussion.
The paper was supported by the Russian
Foundation for Basic Research under grant No. 99-01-00485.

Translated by A. P. Kiselev.

\end{document}